\documentclass[12pt,preprint]{aastex}

\slugcomment{To appear in The Astronomical Journal}

\shorttitle{The Peculiar Radio Source M17 JVLA 35}
\shortauthors{Rodr\'\i guez et al.}

\begin{document}


\title{The Peculiar Radio Source M17 JVLA 35}


\author{L. F. Rodr\'\i guez\altaffilmark{1,2}, C. Carrasco-Gonz\'alez\altaffilmark{1}, G. Montes\altaffilmark{3}}

\and

\author{M. Tapia\altaffilmark{4}}


\altaffiltext{1}{Centro de Radioastronom\'\i a y Astrof\'\i sica UNAM, 
Apartado Postal 3-72 (Xangari), 58089 Morelia, Michoac\'an, Mexico}
\altaffiltext{2}{Astronomy Department, Faculty of Science, King Abdulaziz University, 
P.O. Box 80203, Jeddah 21589, Saudi Arabia}
\altaffiltext{3}{Department of Astronomy and Astrophysics, University of
California, Santa Cruz, CA 95064, USA}
\altaffiltext{4}{Instituto de Astronom\'\i a, Universidad Nacional Aut\'onoma de M\'exico, 
Apdo. Postal 877, Ensenada, BC, CP 22830, Mexico}


\begin{abstract}
M17 JVLA 35 is a radio source detected in projection against
the M17 HII region. In recent observations, its spectrum between 4.96 and 8.46 GHz was found to 
be positive and
very steep, with $\alpha \geq 2.9 \pm 0.6$ ($S_\nu \propto \nu^\alpha$).
Here we present Very Large Array observations made in the 18.5 to 36.5 GHz region
that indicate a spectral turnover at $\sim$13 GHz and
a negative spectral index ($\alpha \simeq -2.0$) at higher frequencies. The spectrum
is consistent with that of an extragalactic High Frequency Peaker (HFP). However,
M17 JVLA 35 has an angular size of $\sim0\rlap.{''}5$ at 8.46 GHz, while HFPs
have extremely compact, milliarcsecond dimensions. 
We discuss other possible models for the spectrum of the
source and do not find them feasible.
Finally, we propose that M17 JVLA35 is indeed
an HFP but that its angular size becomes broadened by plasma scattering as its
radiation travels across M17. 
If our
interpretation is correct, accurate measurements of the angular
size of M17 JVLA35 across the centimeter range should reveal the 
expected $\nu^{-2}$ dependence.

\end{abstract}





\section{Introduction}

In a study made with the Karl G. Jansky Very Large Array (VLA)
at 4.96 and 8.46 GHz toward the classic HII region M17, Rodr\'\i guez et al.
(2012) detected 38 compact radio sources. Of these sources, 19 were found to have
stellar counterparts detected in the infrared, optical, or X-rays.
Among the sources without previously known counterparts, these authors noted that
M17 JVLA 35 (Fig. 1) was remarkable because of its steep positive spectral index
$\alpha \geq 2.9 \pm 0.6$ ($S_\nu \propto \nu^\alpha$).
Sources with such a spectral index at these frequencies are rare. For example, Dzib et al. (2013) determined
the spectral index between 4.5 and 7.5 GHz for 165 sources in the Ophiuchus Complex.
Only two of them have spectral indices comparable to that of JVLA 35.

Counterparts to JVLA 35 were searched unsuccessfully in SIMBAD, as well as in the GLIMPSE catalog 
of candidate young stellar objects with a high probability of association with the M17 complex 
(Povich et al. 2009). Its unusual radio spectrum deserves further study.

In this paper we report new VLA observations of M17 JVLA 35 that extend its spectrum to
the high frequencies: from 18.5 to 36.5 GHz. Our main goal was to obtain a spectrum extending
sufficiently in frequency to attempt a better classification of the source.

\section{Observations}

Observations of M17 were made with the Karl G. Jansky Very Large Array (VLA) of the National 
Radio Astronomy Observatory (NRAO)\footnote{The NRAO is a facility of the National Science 
Foundation operated under cooperative agreement by  Associated Universities, Inc.}. We 
observed the continuum emission in the K and Ka bands in C configuration during June 22nd, 
2013 (project code: 13A-125). For each band, we observed a total continuum bandwidth of 8 GHz 
covering the frequency ranges 18-26 GHz in K band, and 29-37 GHz in Ka band. Each band is 
divided in 4096 channels of 2 MHz each. Bandpass and complex gain calibration were made 
by observing J1825$-$1718 every 2 minutes. Flux calibration was achieved by observation 
of the standard flux calibrator 3C48. We centered our observations at the position of 
M17 JVLA 35 ($\alpha$(J2000)=18$^h$~20$^m$~33.100$^s$, $\beta$(J2000)=$-$16$^\circ$09$'$41.00$''$). 
The on-source time in each band was $\sim$20 minutes. Calibration of the data were performed with 
the data reduction package CASA (Common Astronomy Software Applications; 
version 4.1.0)\footnote{https://science.nrao.edu/facilities/vla/data-processing} 
following standard VLA procedures for high frequency data. 

We divided the calibrated data of each band in 8 parts of 1 GHz each and made cleaned continuum maps of each chunk 
by using the task \emph{clean} of CASA. In order to resolve out the extended emission from the 
M17 HII region, we used only visibilities larger than 50 k$\lambda$ in our imaging, thus 
suppressing structures larger than $\sim$4$"$. A similar imaging procedure was used by Rodriguez et al. (2012).
The central frequencies, rms noises and 
synthesized beams for each map are shown in Table 1. Source JVLA 35 is detected in all our 
1 GHz wide maps and we obtained its flux density at each frequency by performing a Gaussian fit. 
Our typical angular resolution is $\sim1{''}$ and the source is unresolved in our images.

In Table 1 we show the flux density of JVLA 35 obtained from our data and in Figure 2 we show 
the spectral energy distribution (SED). In the SED we have included the flux density at 8.46 GHz 
and the upper limit at 4.96 GHz obtained by Rodr\'{\i}guez et al. (2012). As can be seen, the 
spectral index of the emission is positive at low frequencies, there is a maximum of emission at 
intermediate frequencies, and then the emission diminishes with frequency with a negative spectral 
index at high frequencies.

We fitted the SED by assuming a synchrotron-like spectrum in the form:

\begin{equation}
S_\nu = S_{\nu_0} \left( \frac{\nu}{\nu_0} \right)^{\alpha_{thick}} 
(1 - exp(-\tau_{\nu_0} (\nu/\nu_0)^{\alpha_{thin}-\alpha_{thick}})),
\end{equation}

\noindent where $S_{\nu_0}$ and $\tau_{\nu_0}$ are the flux density and the optical depth at a 
reference frequency $\nu_0$. This equation assumes that the emission is optically thick at low 
frequencies with a spectral index $\alpha_{thick}$, and optically thin at high frequencies 
with a spectral index $\alpha_{thin}$. We assumed the theoretically expected value of $\alpha_{thick}$=2.5 
and obtained $\alpha_{thin}$=$-$2.0 by fitting the data at the highest frequencies. 
The maximum of the emission takes 
place at $\sim$13.3 GHz. 
It should be noted that the spectral index al low frequencies is only
an estimate since we only have an upper limit and one measured flux density.
Additional observations are required. 

\begin{deluxetable}{ccccc}
\tablecolumns{5}
\tablewidth{0pt}
\tablecaption{Flux densities of JVLA 35 at high frequencies}
\tablehead{
\colhead{} Central  &      rms              & \multicolumn{2}{c}{Synthesized beam} & Flux Density   \\
\colhead{} Frequency &     noise             &         HPBW       &       P.A.      & of JVLA 35     \\
\colhead{} (GHz)   & ($\mu$Jy~beam$^{-1}$) &       (arcsec)     &      (deg)      &  ($\mu$Jy)     
}
\startdata
 18.5   &     70                & 1.35$\times$0.74   &       8.90      &  1003 $\pm$ 44 \\
 19.5   &     62                & 1.30$\times$0.72   &       8.15      &  1018 $\pm$ 31 \\
 20.5   &     63                & 1.24$\times$0.68   &       7.95      &   962 $\pm$ 35 \\
 21.5   &     60                & 1.17$\times$0.68   &       9.40      &  1030 $\pm$ 38 \\
 22.5   &     55                & 1.12$\times$0.65   &       9.90      &   838 $\pm$ 64 \\
 23.5   &     55                & 1.09$\times$0.63   &      10.50      &   812 $\pm$ 38 \\
 24.5   &     44                & 1.04$\times$0.60   &      10.98      &   733 $\pm$ 30 \\
 25.5   &     42                & 1.02$\times$0.58   &       8.36      &   647 $\pm$ 33 \\
 29.5   &     68                & 1.14$\times$0.61   &      20.02      &   529 $\pm$ 17 \\
 30.5   &     47                & 1.09$\times$0.59   &      19.83      &   428 $\pm$  7 \\
 31.5   &     46                & 1.02$\times$0.61   &      18.63      &   533 $\pm$ 30 \\
 32.5   &     49                & 0.98$\times$0.60   &      18.85      &   409 $\pm$  6 \\
 33.5   &     44                & 0.97$\times$0.54   &      19.23      &   375 $\pm$ 8  \\
 34.5   &     44                & 0.92$\times$0.58   &      18.25      &   348 $\pm$ 14 \\
 35.5   &     42                & 0.89$\times$0.55   &      19.41      &   328 $\pm$ 16 \\
 36.5   &     44                & 0.88$\times$0.54   &      20.71      &   323 $\pm$ 41 \\
 \enddata
\end{deluxetable}





\section{A Search for a Counterpart in the Infrared}

In order to further constrain the observed properties of M17 JVLA35, we extended our search 
for counterparts to a wider range of  frequencies by analyzing several 
deep survey images  of M17. We retrieved these from from the public archives of the United Kingdom 
Infrared Telescope (UKIRT) Infrared Deep Sky Survey (UKIDSS; Lawrence et al. 2007) in the $JHK$ 
near-infrared bands, the {\sl Spitzer} Galactic Legacy Infrared Midplane Survey Extraordinaire 
(GLIMPSE; Benjamin et al. 2003; Churchwell et al. 2009) taken with the Infrared Array Camera 
(IRAC) at 3.6, 4.5, 5.8 and 8 $\mu$m, and the {\sl Herschel} Infrared Galactic Plane Survey (HI-GAL; 
Molinari et al. 2010) at 70, 160, 250, 350 and 500 $\mu$m. No compact infrared counterparts 
were found in any of these images within $\pm 1''$ of the JVLA position. Using the appropriate 
survey calibrations, we derived the following upper limits to the flux densities: 
16 $\mu$Jy at 1.2 $\mu$m, 23 $\mu$Jy at 1.6 $\mu$m, 17 $\mu$Jy at 2.2 $\mu$m, 
40 $\mu$Jy at 3.6 $\mu$m, 60 $\mu$Jy at 4.5 $\mu$m, 200 $\mu$Jy at 5.8 $\mu$m, 
1 mJy at 8 $\mu$m, 1.2 Jy at 70 $\mu$m and 1.8 Jy at 160 $\mu$m. At 250, 350 and 500
$\mu$m, the extended thermal dust emission associated with M17 is too strong to
derive any useful limit.  As an illustration, Figure 3 shows the $K$-band UKIDDS and
the GLIMPSE 5.8 $\mu$m images of a small field centred on M17 JVLA35.
Note also that no Chandra X-ray source was listed at this position by Broos et al. (2007).

\section{Discussion}

In this section we discuss continuum emission mechanisms
that could explain the spectrum observed for M17 JVLA35.
In addition to the mechanisms discussed below, we note that an
explanation in terms of a galactic background massive star seems
unlikely. These stars have ionized winds that at high frequencies show
a spectral index on +0.6 (i.e. Abbott et al. 1985), in contrast to the negative spectral index
shown by JVLA35.

\subsection{Self-absorbed synchrotron source}

The observed spectrum is similar to those seen in 
the High Frequency Peakers (HFPs).
These are compact, powerful extragalactic
radio sources with well-defined peaks in 
their radio spectra above 5 GHz, with most
of them being high redshift quasars (Dallacasa et al. 2000).
A possible explanation for the HFP radio sources is that we are
observing self-absorbed synchrotron emission. 
An estimate for the magnetic field $B$ in the region can be obtained from
(Kellermann 
\& Pauliny-Toth 1981):

$$\Biggl[{{B} \over {Gauss}} \Biggr] \simeq 9.1 \times 10^{-2}~
\Biggl[{{\nu_{to}} \over {5~GHz}} \Biggr]^5~\Biggl[{{S_{to}} \over {Jy}} \Biggr]^{-2}~
\Biggl[{{\theta_{to}} \over {mas}} \Biggr]^{4},$$

\noindent where $\nu_{to}$, $S_{to}$, and $\theta_{to}$ are the
frequency, flux density, and
angular size of the source at the turnover.
For an HFP radio source, the terms in square brackets on the right hand
side of the equation are of order unity and a magnetic field of order 100 mGauss 
is obtained (e.g. Orienti \& Dallacasa 2014). However, M17 JVLA35 is clearly resolved at
8.46 GHz (with 
deconvolved dimensions of $0\rlap.{''}65\pm0\rlap.{''}09 \times
0\rlap.{''}37\pm0\rlap.{''}06; PA = 19^\circ\pm12^\circ$; Rodr\'\i guez et al. 2012). We will
adopt as the characteristic angular size the geometric mean of the major
and minor axes,
$\sim0\rlap.{''}5$. For this angular dimension the resulting magnetic field is
unrealistically large for any type of astronomical object. We then consider alternative
explanations for the nature of JVLA 35.

\subsection{Optically-thin synchrotron with free-free absorption by M17}

M17 JVLA35 could be a source with a single power-law
spectrum, with the turnover observed at $\sim$13 GHz due to free-free absorption
in the low frequencies
produced by the M17 HII region. 
To test this hypothesis, we calibrated 
VLA archive observations made at 1.42 GHz in the DnC configuration.
These observations were made on 1991 February 12 under project AM279.
We calibrated the data using the usual NRAO procedures and made images
from the average of the pure continuum (that is, without detectable HI absorption) channels.
The data was self-calibrated in phase and the final image is shown in Figure 4.
The total flux density from the HII region in this image
is $\sim$370 Jy. This result indicates that we are detecting most of
the extended emission from the region since the single dish measurement
of Altenhoff et al. (1970) at the same frequency gives a total flux density of 500$\pm$50 Jy.
The position of JVLA 35 is indicated with an $\times$ symbol.

The flux density per beam at the position of JVLA 35 is 2.25 Jy beam$^{-1}$.
In contrast, the flux density per beam at the position of peak emission in the
image ($\alpha = 18^h~ 20^m~ 25\rlap.^s05; \delta = -16^\circ~ 11'~ 34\rlap.{''}7$)
is 9.09 Jy beam$^{-1}$.
These flux densities per beam imply brightess temperatures of $\sim$1,500 K
for the line of sight to JVLA 35 and of $\sim$6,200 K for the position of peak
emission. Assuming that the emission in the latter position traces
optically thick emission, we assume that the electron temperature of
the HII region is $T_e \simeq $ 6,200 K. Then, using

$$T_B = T_e[1 - exp(-\tau_\nu)],$$

\noindent we derive a free-free optical depth of
$\tau(1.42~GHz) \simeq 0.3$ for the emission in the line of sight to JVLA 35. 
Since the free-free optical depth goes as $\nu^{-2.1}$,
we expect $\tau(13~GHz) \simeq 0.002$. We then conclude that the
turnover at 13 GHz cannot be explained by free-free absorption from M17.

\subsection{Optically-thin synchrotron with associated free-free absorption}

Another possibility is an optically-thin synchrotron source that is absorbed at
the lower frequencies by directly associated ionized gas. The problem in this case is that
the optically-thick free-free should produce an emission of about 100 mJy at 8.46 GHz,
while the observed value is of order 1 mJy.

\subsection{Self-absorbed free-free}

The spectral index at low frequencies is marginally
consistent with optically-thick free-free emission ($S_\nu \propto \nu^2$). However,
the clearly negative spectral index above 13 GHz is inconsistent with
optically-thin free-free emission since for this type of emission
the spectral index will always be equal or more positive than $-0.1$ (Rodr\'\i guez et al. 1993).

\subsection{Dust emission}

From the spectrum that rapidly rises in frequency between 4.96 and 8.46
GHz (Rodr\'\i guez et al. 2012), our first assumption was that we
were seeing a source of dust emission. However, dust emission keeps
rising rapidly with frequency ($S_\nu \propto \nu^{2-4}$, e. g.
Rodmann et al. 2006)
well into the far-infrared
and the turnover and change of sign in the spectral index above 13 GHz rules out
this explanation.

\subsection{Spinning dust emission}

Draine \& Lazarian (1998b) proposed that rotating small interstellar grains (spinning dust)
could produce detectable electric dipole radiation. This emission is believed to
account for the "anomalous" Galactic background component, which correlates 
with the 100 $\mu$m thermal emission from dust (Draine \& Lazarian 1998a). 
The spectrum of spinning dust grains is somewhat similar to that
of JVLA 35 in that it rises with frequency, reaches a peak at a
few tens of GHz and drops in intensity for higher frequencies. 
However, the negative spectral index of the high frequency end is 
steeper than the positive spectral index of the low frequency end
(e.g. Ali-Ha{\"i}moud et al. 2009). This behavior is opposite to that
seen in JVLA 35 and we discard this mechanism as a possibility.

\subsection{Self-absorbed synchrotron source with line-of-sight plasma scattering}

As discussed above,
the source M17 JVLA35 should be extremely compact (mas angular size) if it is an
HFP. One possibility is that the source is intrinsically compact, but that in
traveling across M17 the radio emission suffers plasma scattering (e.g. Moran et al. 1990),
a process similar to optical "seeing" that broadens the radio image. A well studied plasma-scattered object is the 
extragalactic source J17204-3554, that is intrinsically very compact but measured to
have an angular dimension of $0\rlap.{''}34 \times
0\rlap.{''}27; PA = 87^\circ$ at 4.9 GHz 
(Trotter et al. 1998). This scattering is produced as the radio waves pass across the
inhomogeneous plasma of the
galactic HII region NGC~6334(A). 

To first approximation, the plasma-scattered angular size is proportional to the
square root of the emission measure along the line of sight to the extragalactic source
(Thompson et al. 1986).
In the case of the extragalactic source J17204-3554, the emission measure of 
NGC~6334(A) is $\sim10^4$ cm$^{-6}$ pc (Moran et al. 1990). For the plasma-scattering
explanation to work, we need a larger emission measure in M17 along the
line-of-sight to the presumed extragalactic HFP.
Since the angular size of the scattered source goes as $\nu^{-2}$ and M17 JVLA35 at 8.46 GHz is about twice
as large as J17204-3554 at 4.9 GHz, we need that the emission measure in the line-of-sight
within M17 is about 40 times that of NGC~6334(A).

As noted previously, we derive an optical depth of
$\tau(1.42~GHz) \simeq 0.3$ for the emission in the line of sight to 
JVLA 35.  This optical depth corresponds to an emission
measure of $\sim10^6$ cm$^{-6}$ pc.
This emission measure is about
100 times larger than that measured for NGC~6334(A) in the line of sight to
J17204-3554. We then conclude that M17
can easily provide the observed plasma scattering.
We note that the emission measure toward
the region of peak emission of M17 is even larger,
$\geq4 \times 10^6$ cm$^{-6}$ pc)

A direct way to test this hypothesis is to determine the angular size of M17 JVLA 35 at
several frequencies, searching for the expected $\nu^{-2}$ dependence.
At present, we cannot perform this test because the source was detected by Rodr\'\i guez et al.
(2012) only at one frequency and because the observations reported here lack
the angular resolution to achieve this. Additional observations are required to
more fully understand the nature of M17 JVLA 35. One expects to detect additional compact
extragalactic sources that are plasma-scattered if they are located behind a galactic HII region or
planetary nebula.

\section{Conclusions}

We discussed several possibilities to explain the observed spectrum and
angular size of the source M17 JVLA35. 
We favor that it is an HFP whose angular size is plasma-broadened as the radiation
travels across M17. This hypothesis can be tested by searching the expected
$\nu^{-2}$ angular size dependence in the centimeter range.







\acknowledgments

This research has made use of the SIMBAD database,
operated at CDS, Strasbourg, France.
CCG and LFR are grateful to CONACyT, Mexico and DGAPA, UNAM for their financial
support.
The UKIDSS project is defined in Lawrence et al (2007). UKIDSS uses the UKIRT Wide Field Camera 
(WFCAM; Casali et al (2007) and a photometric system described in Hewett et al. (2006). 
The pipeline processing and science archive are described in Hambly et al (2008). 
The {\sl Spitzer Space Telescope}, which is operated by the Jet Propulsion Laboratory, 
California Institute of Technology (CIT)  under National Aeronautics and Space 
Administration (NASA) contract 1407.  {\sl Herschel} is an ESA space observatory with science
instruments provided by European-led Principal Investigator consortia and with important 
participation from NASA.



{\it Facilities:} \facility{VLA, UKIDDS, GLIMPSE and HI-GAL}.

\clearpage



\begin{figure}
\epsscale{1.0}
\plotone{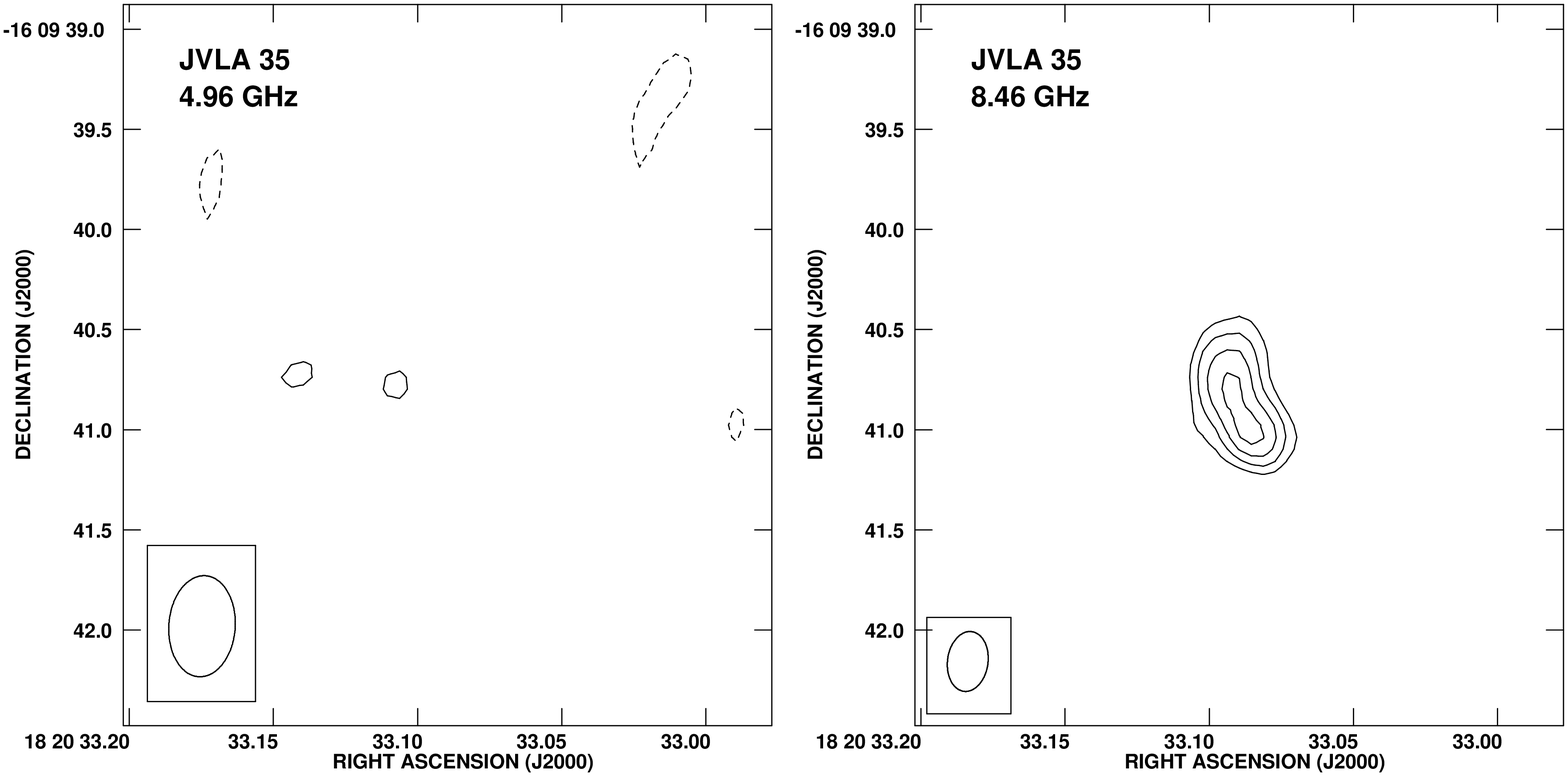}
\caption{JVLA images of the source JVLA 35 at 4.96 (left) and 8.46 GHz (right)
from Rodr\'\i guez et al. (2012).
Contours are $-3$, 3, 4, 5, and 6 times 30$\mu$Jy beam$^{-1}$.
The source is clearly detected at 8.46 GHz, but not at 4.96 GHz.
The half power contour of the synthesized beam is shown in the
bottom left corner of the images and is
$0\rlap.{''}51 \times 0\rlap.{''}33$ with
$PA = -4^\circ$
for the
4.96 GHz image and 
$0\rlap.{''}30 \times 0\rlap.{''}20$ with
$PA = -8^\circ$ for the
8.46 GHz image.}
\end{figure}
\clearpage		      

\begin{figure}
\epsscale{0.8}
\plotone{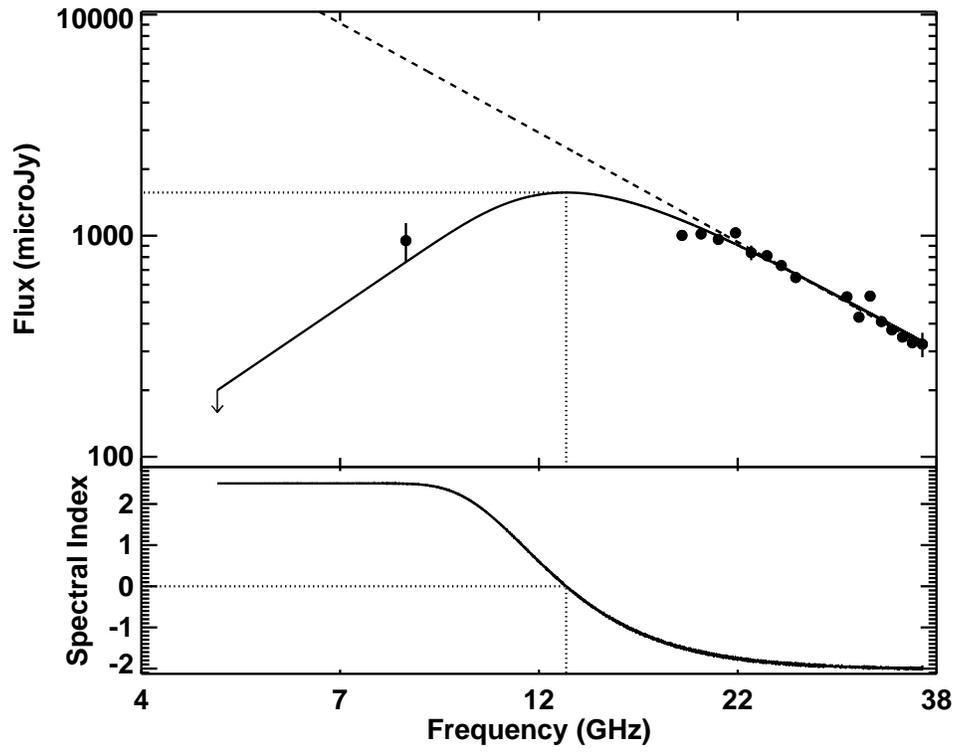}
\caption{(Top) Spectral energy distribution (SED) in the radio for M17 JVLA35.
(Bottom) Spectral index across the SED.}
\end{figure}

\clearpage

\begin{figure*}
\center
\resizebox{10cm}{!}{\includegraphics[clip=true]{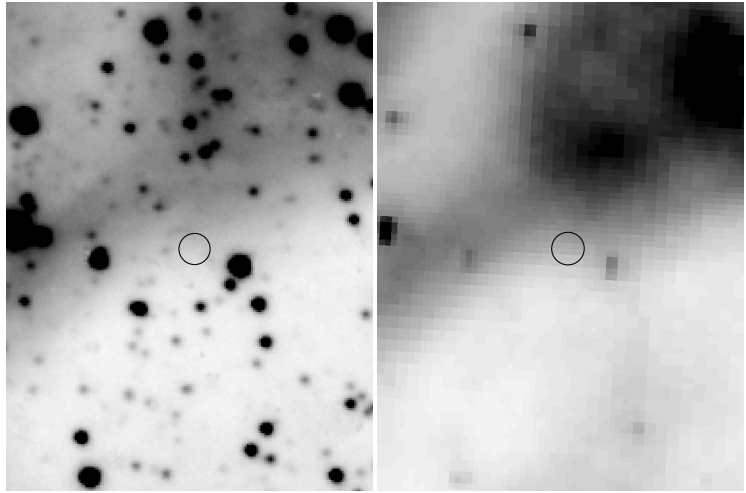}}
\caption{Near- and mid-infrared images of a $37'' \times 50''$ field centered at the JVLA position 
of M17~JVLA35, indicated by a circle of $3''$ diameter. The left panel is the $2 \times 2$ kernel
smoothed UKIDDS $K_s$ band image and the right panel is the $Spitzer$/IRAC 5.8 $\mu$m 
image.}
\label{fig3}
\end{figure*}

\clearpage

\begin{figure}
\epsscale{.80}
\plotone{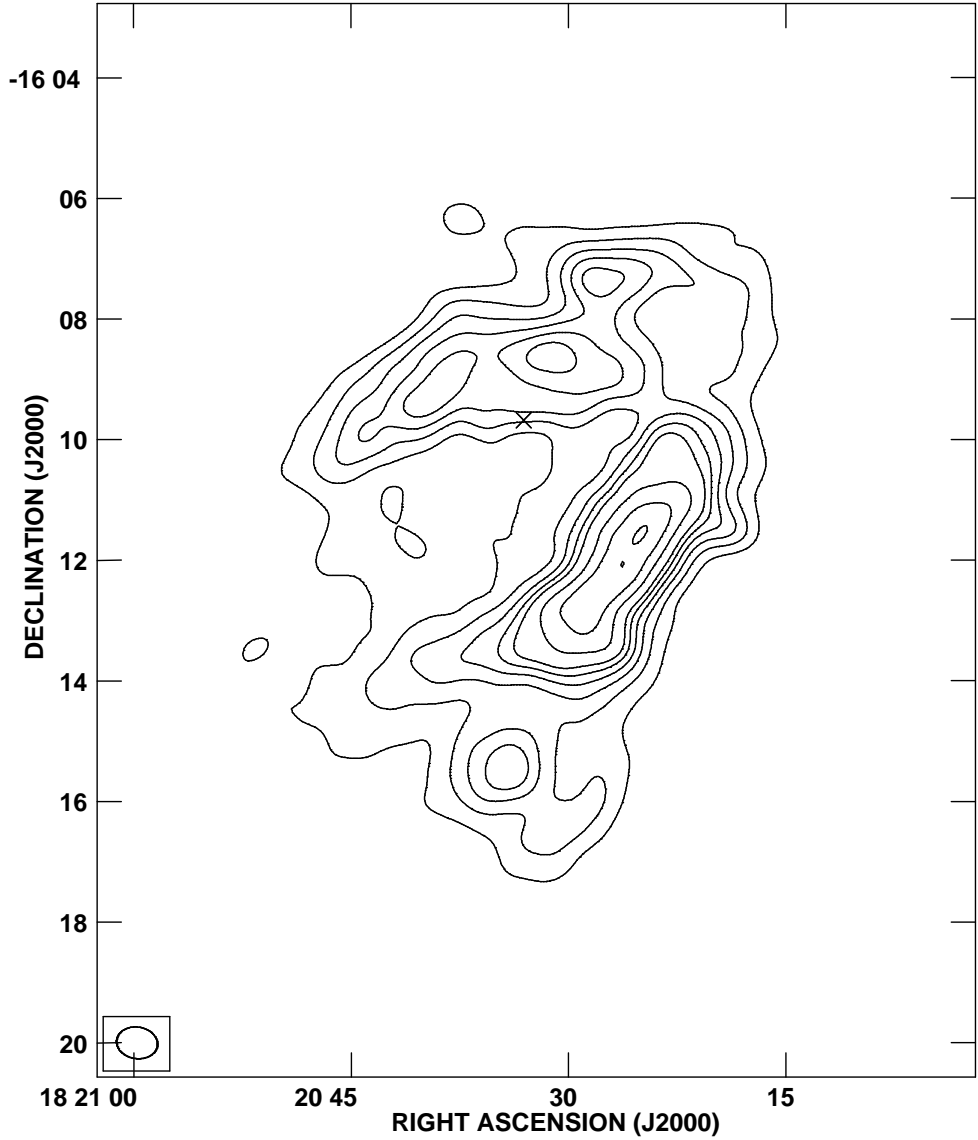}
\caption{VLA image of M17 at 1.42 GHz. Contours are
-5, 5, 10, 15, 20, 25, 30, 40, 50 and 60 times 0.15 Jy beam$^{-1}$.
The position of VLA 35 is indicated with an $\times$ symbol.
The synthesized beam ($41\rlap.{''}0 \times 31\rlap.{''}4; PA = +82^\circ$) is
shown in the bottom left corner. The image has been corrected for the primary
beam response.}
\end{figure}

\clearpage

\clearpage


\end{document}